\documentclass[english,aps,prl, twocolumn,superscriptaddress]{revtex4-1}
\usepackage[utf8]{inputenc}
\usepackage{amsmath}
\usepackage{amssymb}
\usepackage{graphicx}
\usepackage{hyperref}
\usepackage{siunitx}
\usepackage{placeins}
\usepackage{braket}
\usepackage{babel}
\usepackage{mathtools}
\usepackage{xspace}
\usepackage{xcolor}
\usepackage{physics}

\newcommand{\comment}[1]{}

\begin{document}
\title{Hilbert-space geometry of random-matrix eigenstates}
\author{Alexander-Georg Penner}
\affiliation{\mbox{Dahlem Center for Complex Quantum Systems and Fachbereich Physik, Freie Universit\"at Berlin, 14195, Berlin, Germany}}
\author{Felix von Oppen}
\affiliation{\mbox{Dahlem Center for Complex Quantum Systems and Fachbereich Physik, Freie Universit\"at Berlin, 14195, Berlin, 
Germany}}
\author{Gergely Zar\'and}
\affiliation{Exotic Quantum Phases ``Momentum'' Research Group, Department of Theoretical Physics, Budapest University of Technology and Economics, 1111 Budapest, Budafoki út 8, Hungary}
\affiliation{MTA-BME Quantum Correlations Group, Institute of Physics, Budapest University of Technology and Economics, 1111 Budapest, Budafoki út 8, Hungary}
\author{Martin R. Zirnbauer}
\affiliation{Institut f\"ur Theoretische Physik, Universit\"at zu K\"oln, Z\"ulpicher Straße 77a, 50937 K\"oln, Germany}

\begin{abstract}
The geometry of multi-parameter families of quantum states is important in numerous contexts, including adiabatic or nonadiabatic quantum dynamics, quantum quenches, and the characterization of quantum critical points. Here, we discuss the Hilbert-space geometry of eigenstates of parameter-dependent random-matrix ensembles, deriving the full probability distribution of the quantum geometric tensor for the Gaussian Unitary Ensemble. Our analytical results give the exact joint distribution function of the Fubini-Study metric and the Berry curvature. We discuss relations to Levy stable distributions and compare our results to numerical simulations of random-matrix ensembles as well as electrons in a random magnetic field. 
\end{abstract}

\maketitle

{\em Introduction.---}The geometry underlying the eigenstates of parameter-dependent quantum Hamiltonians is concisely described in terms of the quantum geometric tensor \cite{ProvostVallee1980,VenutiZanardi2007}. Its symmetric part is the Fubini-Study metric, while its antisymmetric part is the Berry curvature \cite{Berry1984}. Both contributions to the quantum geometric tensor have important physical consequences, in particular in the context of adiabatic quantum dynamics beyond the Born-Oppenheimer approximation. When a slow system is coupled to a fast one, the symmetric and antisymmetric parts of the quantum geometric tensor govern electric and magnetic gauge forces acting on the slow system. An important application of these ideas is to the semiclassical dynamics of Bloch electrons \cite{Xiao2010}, where these gauge forces are at the core of anomalous Hall effects, both unquantized and quantized. In this case, each band defines a family of quantum states which is parametrized by the Bloch momenta, and it is by now well understood that the physics of electronic systems is affected by the local geometry \cite{Xiao2010} as well as the global topology of the bands \cite{Qi2011}. In disordered or interacting systems, the magnetic fluxes threading the system in a real-space torus geometry play a role which is quite analogous to that of the Bloch momenta of noninteracting clean systems \cite{Niu1985}. The corresponding boundary geometric tensor has been shown to provide an appropriate scaling variable for Anderson transitions, and to assume a universal probability distribution at the critical point \cite{Werner2019}. More generally, the quantum geometric tensor is an important characteristic of quantum phase transitions \cite{VenutiZanardi2007,Carollo2020}.

Here, we derive the exact joint probability distribution of the quantum geometric tensor for the Gaussian Unitary Ensemble (GUE) of random-matrix theory. The probability distribution of the quantum geometric tensor for random-matrix ensembles was recently introduced by Berry and Shukla \cite{Berry2020}, extending earlier work on the Berry curvature \cite{Simons1998,Shukla2018,Shukla2019}. Berry and Shukla base their discussion on analytical results for small random matrices, which is sufficient to obtain the correct asymptotics of the distribution function, but fails to describe the bulk of the distribution for generic systems. Here, we find the exact analytical distribution in the limit of large random matrices. Large random matrices are a powerful tool to describe spectra and eigenstates of generic quantum systems \cite{Dyson1962,Guhr1998} and are applicable to a remarkably diverse set of systems, including nuclear spectra \cite{Brody1981}, quantum chromodynamics \cite{Verbaarschot2000}, few-body chaotic quantum systems \cite{Bohigas1984}, disordered electron systems \cite{Efetov1983,Beenakker1997}, nonintegrable many-body systems \cite{Poilblanc1993,Santos2010}, and many-body localization \cite{Pal2010,Serbyn2016,Filippone2016}. Most recently, random-matrix theory was instrumental in claims that quantum processors have reached the regime of quantum supremacy \cite{Arute2019}. A central role in this argument was played by the Porter-Thomas distribution, one of only few distribution functions in random-matrix theory which are known exactly and have a simple analytical form. In view of the scarcity of exact analytical distributions in random-matrix theory, it is quite remarkable that the characteristic function of the joint distribution function of the quantum geometric tensor can be obtained exactly. 

{\em Quantum geometric tensor.---}We consider the eigenstates $|\tilde n({\boldsymbol {\lambda}})\rangle$ of a multi-parameter family of Hamiltonians $H({\boldsymbol {\lambda}})$ with ${\boldsymbol {\lambda}}=(\lambda_1,\ldots,\lambda_n)$. A metric structure associated with the parameter-dependent eigenstates can be obtained by defining the distance in Hilbert space for two states with infinitesimally different parameters as
\begin{equation}
   ds^2 = 1 - |\langle \tilde n({\boldsymbol {\lambda}})|\tilde n({\boldsymbol {\lambda}}+d{\boldsymbol {\lambda}})\rangle|^2 = \sum_{\alpha\beta} \textrm{Re} g^{(n)}_{\alpha\beta}({\boldsymbol {\lambda}}) d\lambda_\alpha d\lambda_\beta .
\label{metric}
\end{equation}
Explicitly expanding in $d{\boldsymbol {\lambda}}$ yields the Hermitian quantum geometric tensor \cite{ProvostVallee1980,VenutiZanardi2007}
\begin{equation}
   g_{\alpha\beta}^{(n)} = \langle \partial_\alpha \tilde n|\partial_\beta \tilde n\rangle  - \langle \partial_\alpha \tilde n|\tilde n\rangle  \langle \tilde n | \partial_\beta \tilde n\rangle.
\label{QGT}
\end{equation}
The distance $ds^2$ is entirely determined by the real and symmetric part, which is also known as the quantum metric tensor. The imaginary and antisymmetric part is readily identified as the Berry curvature \cite{Berry1984}, which can be nonzero for broken time-reversal symmetry. Equation (\ref{metric}) indicates that the quantum geometric tensor $g_{\alpha\beta}^{(n)}$ quite generally governs the behavior of systems under quantum quenches which involve small changes of the parameters. 

Following Berry and Shukla \cite{Berry2020}, we consider a two-parameter family of Hermitian $N\times N$ Hamiltonians 
\begin{equation}
    H = H_0 + x H_x + y H_y ,
\label{ham}
\end{equation}
which depend on the real parameters $x$ and $y$. Evaluating the derivatives in Eq.\ (\ref{QGT})  at $x=y=0$, one can express
the quantum geometric tensor in terms of the eigenenergies $E_n$ and eigenstates $|n\rangle$ of $H_0$,
\begin{equation}
    g_{\alpha\beta}^{(n)} = \sum_{m(\neq n)} \frac{\langle n| H_\alpha|m\rangle \langle m| H_\beta|n\rangle}{(E_n-E_m)^2}
\label{eq:Cn}
\end{equation}
with $\alpha,\beta \in \{x,y\}$.

For orientation, we first consider the distribution function of individual matrix elements of the quantum geometric tensor for an $N\times N$ matrix Hamiltonian $H_0$ of an integrable system whose energy eigenvalues are statistically independent. In this case, the matrix elements of the quantum geometric tensor in Eq.\ (\ref{eq:Cn}) are sums over $N-1$ statistically independent terms, $g_{\alpha\beta}^{(n)} = \sum_{m(\neq n)} x_m$, and one expects their probability distributions $P_\mathrm{int}(g)$ to converge to a stable distribution in the limit $N\to\infty$. In the absence of correlations between the eigenvalues and thus of level repulsion, the distribution of the individual terms in the sum is readily seen to fall off as $1/|x|^{3/2}$ at large $|x|$ \cite{supp}, with large values of $|x|$ originating from near degeneracies in the spectrum of $H_0$. Importantly, both the average and the variance diverge for this distribution. As a result, the sum (\ref{eq:Cn}) does not constitute a standard random walk, for which the central limit theorem predicts a normal distribution. Instead, the matrix elements $g_{\alpha\beta}^{(n)}$ can be viewed as Levy flights and their probability distributions are Levy stable distributions. The terms in the sum have random signs for the real and imaginary parts of off-diagonal matrix elements, but are strictly positive for diagonal elements, leading to different stable distributions. For an asymptotic $1/|x|^{3/2}$ decay at large $|x|$, one finds distributions $P_\mathrm{int}(g) = \int\frac{\mathrm{d}\xi}{2\pi} e^{i\xi g} \tilde P_\mathrm{int}(\xi)$ with characteristic functions \cite{Bouchaud1990,supp}
\begin{equation}
    \tilde P_\mathrm{int}(\xi) = \left\{ \begin{array}{ccc} e^{-\sqrt{\frac{1}{2}\gamma |\xi|}(1+i\,\textrm{sgn}\xi)} &\,\,\,\,\,\,\, & \textrm{diagonal} \\  
     e^{-  \sqrt{\gamma |\xi|}} && \textrm{off-diagonal} \end{array} \right. ,
\label{levy}
\end{equation}   
where $\gamma$ controls the scale. Due to the $\sqrt{|\xi|}$ singularity of the characteristic function, the distributions $P_\mathrm{int}(g)$ fall off as $1/|g|^{3/2}$ at large $|g|$, indicating that they are dominated by individual terms in the sum (\ref{eq:Cn}). Physically, this broad distribution is a direct consequence of the fact that the level spacing distribution of integrable systems remains nonzero in the limit of zero spacing. 

{\em Joint distribution function for the GUE.---}In generic systems, level repulsion suppresses the likelihood of small energy denominators and the distribution of matrix elements of the quantum geometric tensor decays faster. If we continue to assume that the matrix elements are dominated by individual terms in the sum (\ref{eq:Cn}), the tail of the distribution can be predicted on the basis of random $2\times2$ GUE matrices, yielding a faster asymptotic decay, $\sim 1/|g|^{5/2}$ \cite{Berry2020}. In addition to suppressing the probability with which near degeneracies occur, level repulsion introduces correlations between the terms in the sum in Eq.\ (\ref{eq:Cn}).  As a result, the distribution of the quantum geometric tensor no longer belongs to the family of Levy stable distributions. Remarkably, however, it can still be computed exactly. 

We now focus on large random matrices drawn from the Gaussian Unitary Ensemble, which neither obeys time-reversal symmetry nor imposes any other (anti)symmetry (symmetry class A in the Altland-Zirnbauer classification \cite{Zirnbauer1996,Altland1997}). The (Hermitian) matrices $H_0$, $H_x$, and $H_y$ are drawn from three statistically independent GUEs, 
\begin{eqnarray}
   &&P(H_0,H_x,H_y) {\mathrm d}H_0{\mathrm d}H_x{\mathrm d}H_y  \nonumber\\
   &&\,\,\,\,\,\, \propto e ^{-\frac{1}{2}N \mathrm{tr} (H_0^2 + H_x^2 + H_y^2) }  \prod_{i,j}{ d}(H_0)_{ij}{ d}(H_x)_{ij}{ d}(H_y)_{ij}, \,\,\,\,
\label{GUE}
\end{eqnarray}
where the averages over $H_x$ and $H_y$ are introduced for convenience. We comment below on the case when the average is over $H_0$ only. Exploiting Hermiticity, we parametrize the quantum geometric tensor $g^{(n)}$ as $g^{(n)}=g_0^{(n)}+\mathbf{g}^{(n)}\cdot\boldsymbol{\tau}$, where $\boldsymbol{\tau}$ denotes the vector of Pauli matrices and $g_0^{(n)}=\frac{1}{2}\mathrm{tr} g^{(n)}$, $g_1^{(n)}= \mathrm{Re}g_{yx}^{(n)}$, $g_2^{(n)}= \mathrm{Im}g_{yx}^{(n)}$, and $g_3^{(n)} = \frac{1}{2}\mathrm{tr} (\tau_3 g^{(n)})$. Notice that $g_2^{(n)}$ measures the Berry curvature and $g_0^{(n)}, g_1^{(n)}, g_3^{(n)}$ parametrize the quantum metric tensor. We can then define the joint probability distribution of the quantum geometric tensor through
\begin{equation}
    P(g) \propto \left\langle \sum_n \delta(E_n) \delta(g_0 - g_0^{(n)}) \delta(\mathbf{g} - \mathbf{g}^{(n)}) \right\rangle_{H_0,H_x,H_y}.
\end{equation}
Here, the brackets denote the random-matrix average and the first $\delta$-function ensures that we consider the quantum geometric tensor for states which are at the center of the spectrum~\footnote{We systematically ignore prefactors which can be restored from the normalization condition at the end of the calculation.}. 

The corresponding characteristic function defined via $P(g)=\int \frac{d\xi_0}{2\pi} \frac{d\boldsymbol{\xi}}{(2\pi)^3} e^{i [\xi_0 g_0 + \boldsymbol{\xi}\cdot\mathbf{g}]} \tilde P(\xi_0,\boldsymbol{\xi})$ takes the form
\begin{equation}
   \tilde P(\xi_0,\boldsymbol{\xi}) \propto \left\langle \sum_n \delta(E_n) e^{-i [\xi_0 g_0^{(n)} + \boldsymbol{\xi}\cdot\mathbf{g}^{(n)}]}\right\rangle_{H_0,H_x,H_y}.
\label{Palpha}
\end{equation}
In the limit of $N\to\infty$, the random-matrix averages can be performed explicitly. We defer technical details to further below and the supplemental material \cite{supp}, and focus first on discussing our results. 

{\em Results.---}We find that the characteristic function for the quantum geometric tensor takes the exact form
\begin{equation}
  \tilde{P}(\xi_0,\boldsymbol{\xi}) = r(X_+, X_-) 
    e^{-(X_++X_-) },
   \label{final1}
\end{equation}
where we defined $X_\pm = \tfrac{1 }{2}(1+i\mathrm{sgn}\xi_\pm )\sqrt{\gamma |\xi_\pm|}$ in terms of 
$\xi_\pm= \xi_0\pm |\boldsymbol{\xi}|$ and the rational function
\begin{widetext}
\begin{eqnarray}
   r(a,b) &=& 1 + (a+b) +\frac{1}{3}(a^2+3ab+b^2) + \frac{1}{24}\frac{a^4+9a^3b + 17a^2 b^2 + 9 ab^3 + b^4}{a+b}
+\frac{1}{120}ab(5a^2+16ab+5b^2)
\nonumber\\
  &&  +\frac{1}{720}\frac{a^2b^2(13a^2+29ab+13b^2)}{a+b}
+\frac{1}{240}a^3b^3 + \frac{1}{1920} \frac{a^4b^4}{a+b} + \frac{1}{34560} \frac{a^5b^5}{(a+b)^2}.
\label{final2}
\end{eqnarray}
\end{widetext}
For the specific scalings of the GUE matrices in Eq.\ (\ref{GUE}), we find $\gamma^{\textrm{GUE}}=4{N}$. Notice that $\tilde P(0,\mathbf{0})=1$, so that $P(g)$ is normalized. Equations~(\ref{final1}) and (\ref{final2}) give the exact characteristic function of the distribution of the quantum geometric tensor for large GUE matrices, and are the central results of this paper. 

We first specify Eqs.\ (\ref{final1}) and (\ref{final2}) to the distribution of individual matrix elements of $g$. The characteristic function of the distribution of the diagonal elements $g_{xx}$ and $g_{yy}$ can be obtained by setting $\xi_0=\pm\xi_3=\xi$ and $\xi_1=\xi_2=0$. Interestingly, the resulting exponential factor in Eq.\ (\ref{final1}) has just the same form as in Eq.\ (\ref{levy}). The same happens for the distributions of $\mathrm{Re}g_{xy}$ and the Berry curvature $\mathrm{Im}g_{xy}$, which are obtained from Eq.\ (\ref{final1}) by setting $\xi_1=\xi$ or $\xi_2=\xi$, respectively, with all other $\xi_j=0$. Thus, it is the rational prefactor in Eq.\ (\ref{final1}) that accounts for the spectral correlations introduced by the GUE. Expanding the exponential in Eq.\ (\ref{final1}), we observe that the leading nonanalyticity of $\tilde P(\xi_0,\boldsymbol{\xi})$ is of the form $|\xi|^{3/2}$, which contrasts with the leading $|\xi|^{1/2}$ singularity of the characteristic function $\tilde P_\mathrm{int}(\xi)$ in Eq.\ (\ref{levy}). This implies that for the GUE, the distribution function of the quantum geometric tensor indeed falls off as $P(g)\propto 1/|g|^{5/2}$ for large $|g|$ and thus faster than the corresponding distribution $P_\mathrm{int}(g)\propto 1/|g|^{3/2}$ for integrable systems, corroborating the expectation based on $2\times 2$ GUE matrices \cite{Berry2020}. 

\begin{figure}[b]
	\includegraphics[width=0.49\linewidth]{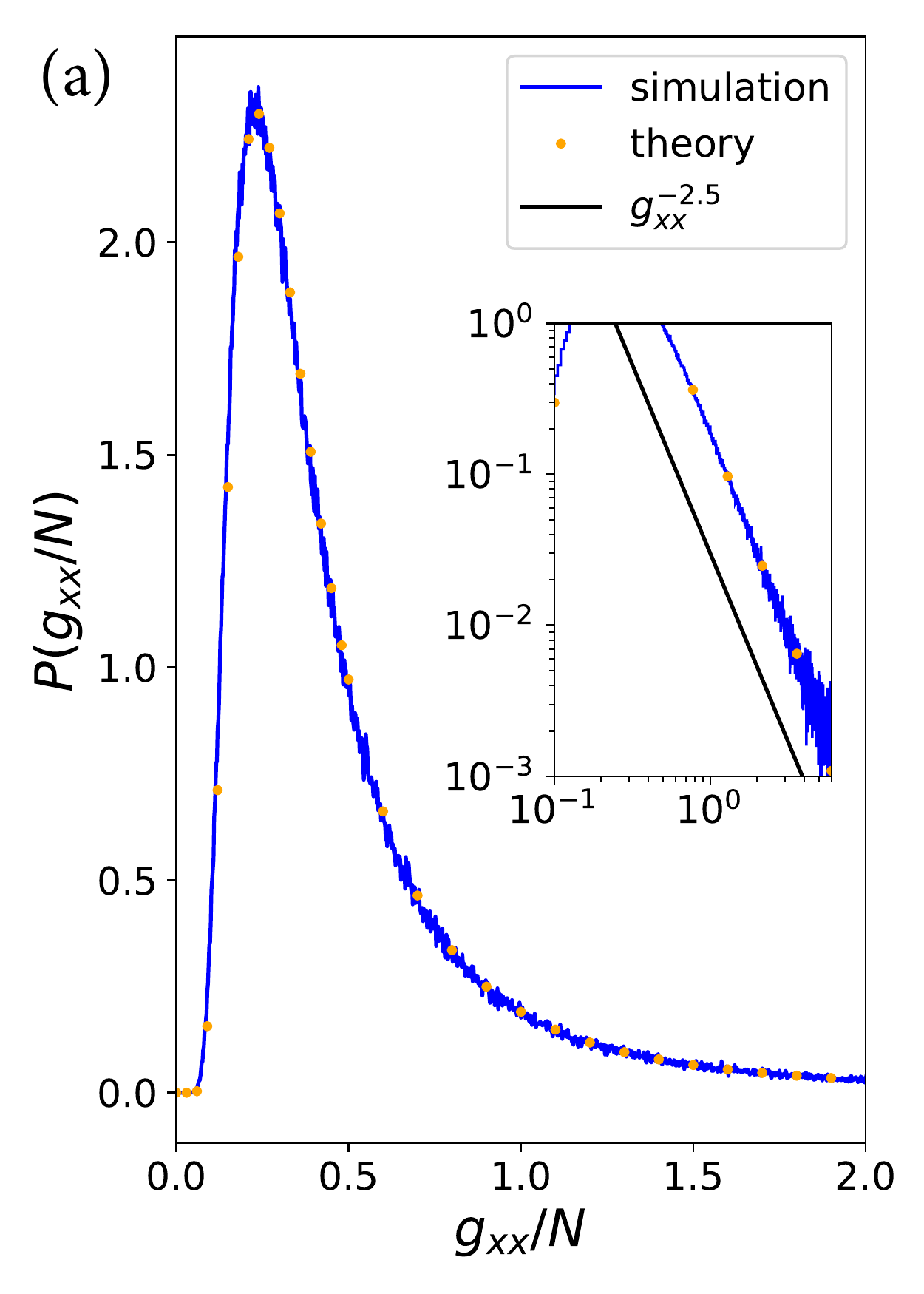} 
	\includegraphics[width=0.49\linewidth]{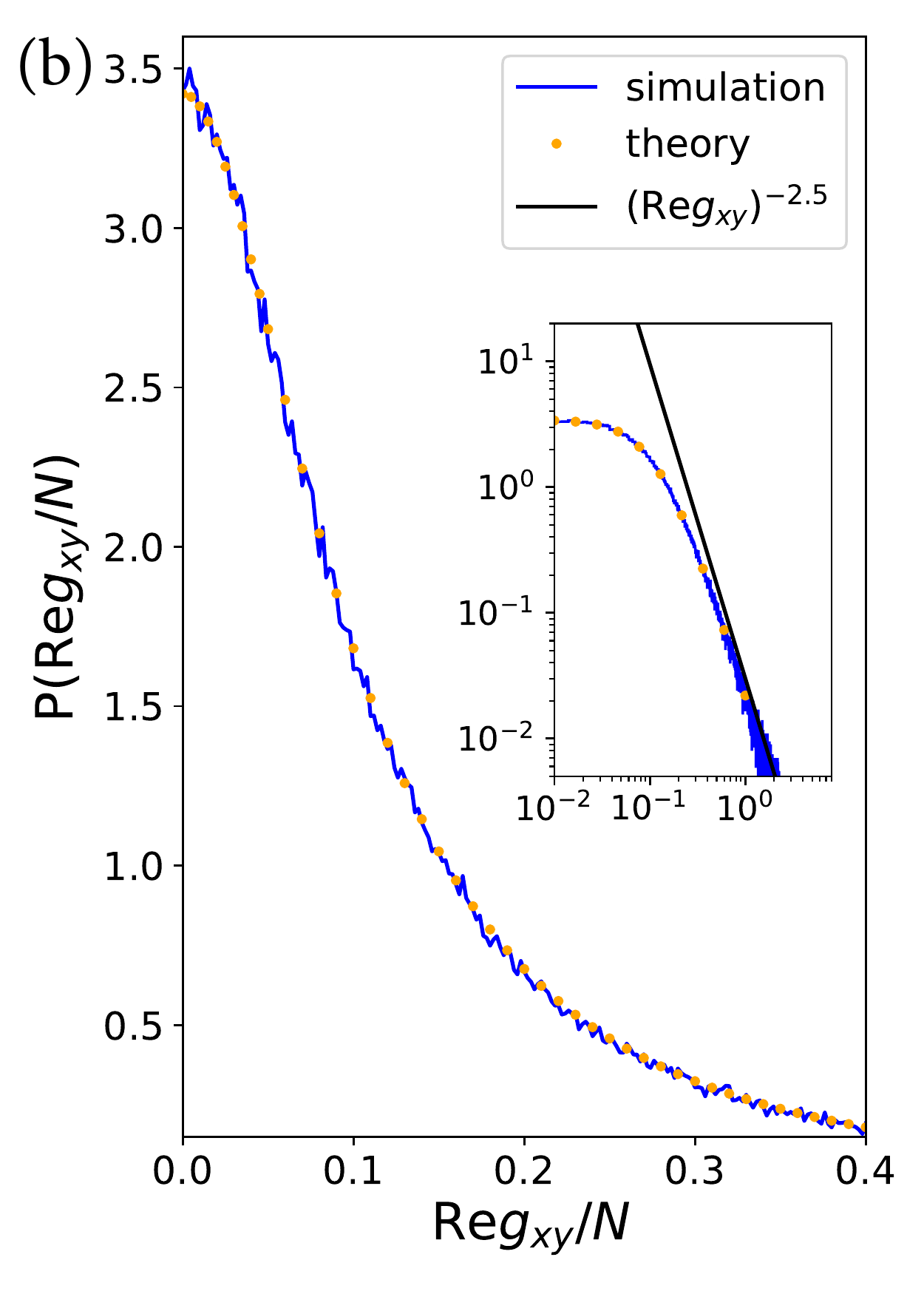}
	\includegraphics[width=0.49\linewidth]{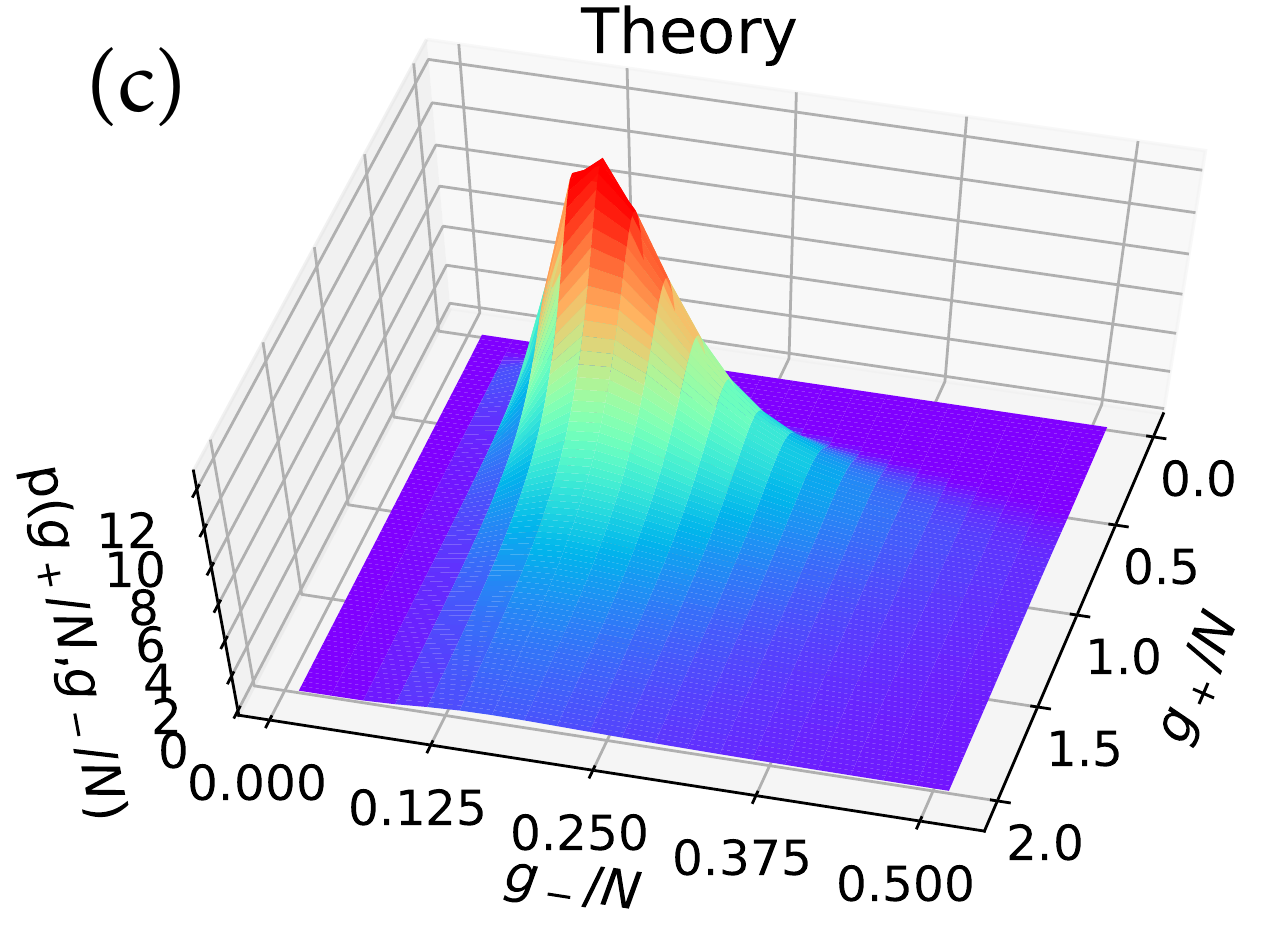} 
	\includegraphics[width=0.49\linewidth]{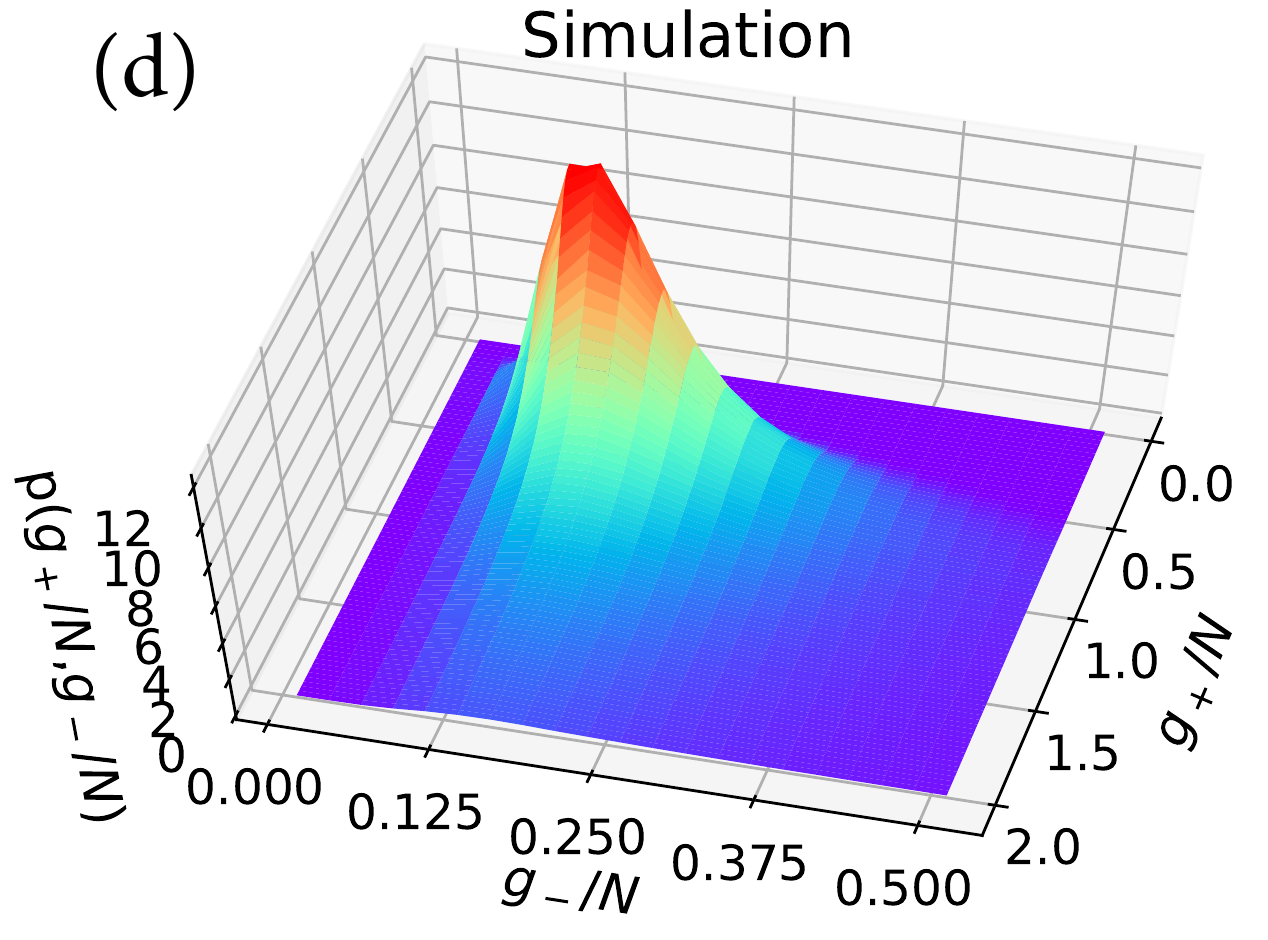} 
		 \caption{\label{Fig1} Distribution functions of (a) the diagonal and (b) the off-diagonal matrix element (real part) of the quantum geometric tensor. Numerical data for large random matrices (blue lines) are compared to the Fourier transform of the analytical result obtained from Eq.\ (\ref{final1}) (orange dots). (c) 3D plot of the distribution function $p(g_+,g_-)$ based on the analytical result in Eq.\ (\ref{gplusgminus}). (d) Corresponding 3D plot obtained numerically for large random matrices, obtained by averaging over $10^6$ realizations of $H$ in Eq.\ (\ref{ham}) with $H_0$, $H_x$, and $H_y$ drawn independently from the GUE  with $N=100$. Insets in (a) and (b): Log-log plots emphasizing the asymptotic $1/|g|^{5/2}$ decays (black line).}
\end{figure} 

Our analytical distribution functions of the diagonal and off-diagonal components of the quantum metric tensor are in excellent agreement with numerical results for GUE random matrices as shown in Figs.\ \ref{Fig1}(a) and (b). Moreover, we find that the off-diagonal element of the quantum metric tensor has the same distribution as the Berry curvature \cite{Simons1998}. To compare our analytical results to numerical simulations for GUE random matrices in more detail, we note that $P(g)$ obtained by Fourier transforming Eq.\ (\ref{final1}) depends on the quantum geometric tensor only through its eigenvalues $g_\pm = g_0\pm |\mathbf{g}|$. Writing $g=U\mathrm{diag}[g_+,g_-]U^\dagger$, the distribution function is independent of the diagonalizing unitary matrix $U$, and employing a convenient redundancy of parametrization, we define the corresponding joint eigenvalue distribution $p(g_+,g_-)$ through  
\begin{equation}
    P(g) dg = p(g_+,g_-) dg_+dg_-d\mu(U), 
\end{equation}
where $d\mu(U)$ is the invariant measure of the unitary group, with the group volume normalized to unity. We find
\begin{eqnarray}
   p(g_+,g_-) &=& -\frac{i(g_+-g_-)}{32\pi^2} \int d\xi_+ d\xi_- (\xi_+-\xi_-) 
    \nonumber\\    
   && \,\,\,\,\,\, \times \tilde P(\xi_0,\boldsymbol{\xi}) 
    e^{\frac{i}{2}(g_+\xi_+ + g_-\xi_-)}
\label{gplusgminus}
\end{eqnarray} 
A 3D plot of this distribution is shown in Fig.\ \ref{Fig1}(c) and compared to a numerical histogram for GUE matrices in Fig.\ \ref{Fig1}(d), again showing excellent agreement. 

The GUE averages over the perturbations $H_x$ and $H_y$ are actually redundant in the limit of $N\to \infty$ considered above. In \cite{supp}, we show both analytically and numerically that one obtains the same distribution (\ref{final1}) when averaging over the unperturbed GUE Hamiltonian $H_0$ only. 
 
\begin{figure}[t]
\includegraphics[width=0.97\linewidth]{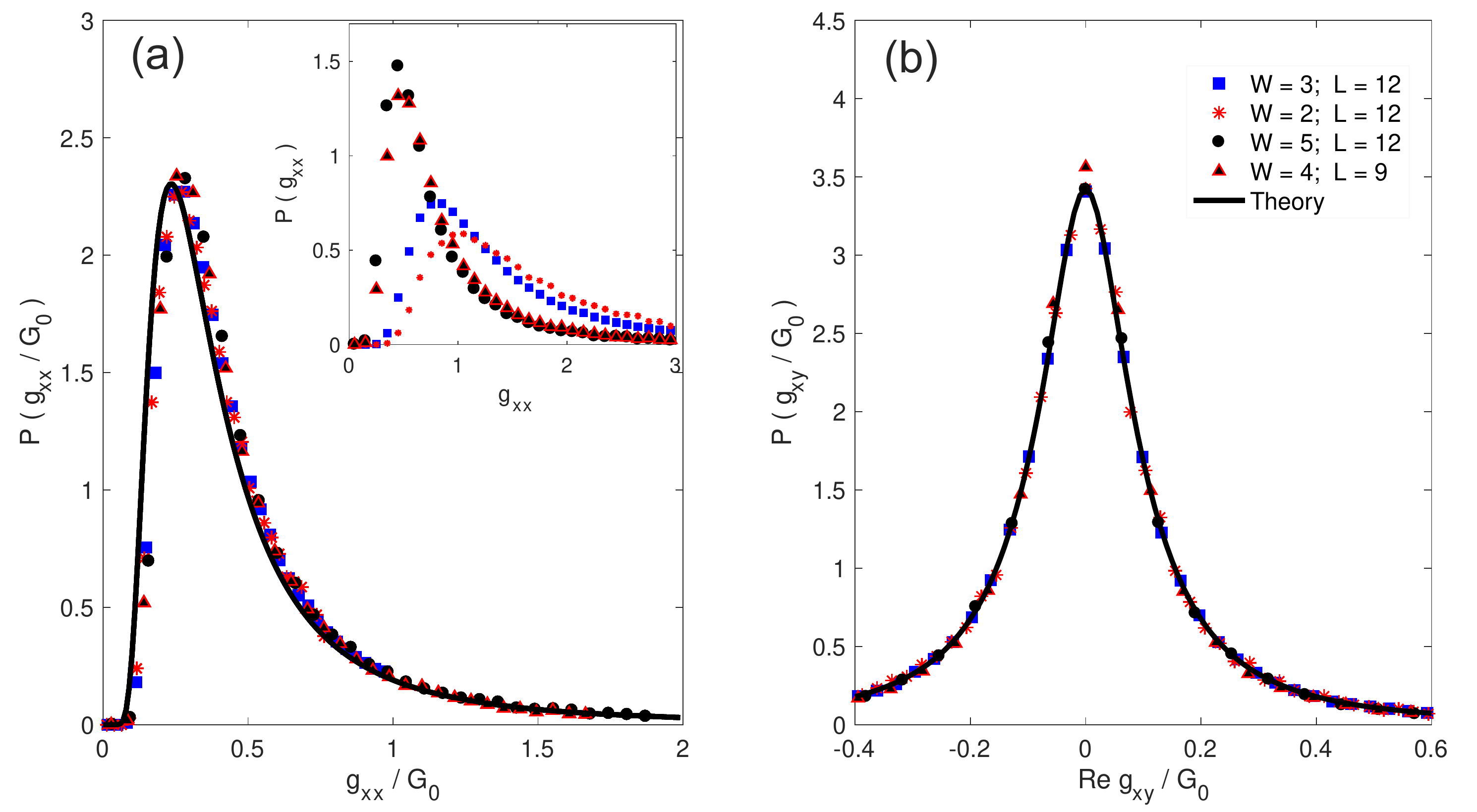} 
\caption{\label{Fig2} Distribution functions of (a) the diagonal and (b) the off-diagonal matrix element (real part) of the quantum geometric tensor of the 3D random flux model. Numerical data [symbols; see legend in panel (b)] are compared to the analytical result obtained from Eq.\ (\ref{final1}) (full line). The inset in panel (a) shows unscaled data for $g_{xx}$. The data in the main panels were scaled to collapse onto a universal curve using the same set of scaling factors $G_0$ for $g_{xx}$ in (a) and ${\rm Re}[g_{xy}]$ in (b), namely $G_0=3.0395$ for $W = 3;  L = 12$, $G_0=3.8685$ for $W = 2;  L = 12$, $G_0=1.5760$ for $W = 5;  L = 12$, and $G_0=1.7730$ for $W = 4;  L = 9$.}
\end{figure}

{\em Random-flux model}.---The distribution function of the quantum geometric tensor is thus not very sensitive to the particular nature of the pertubation. This suggests that it applies to the large class of physical models which have been shown to display GUE random-matrix correlations. Here, we illustrate this broad applicability by simulations for an appropriate Anderson model. Specifically, we consider a tight-binding model 
\begin{equation}
   H = \sum_{\langle ij \rangle} t^{\phantom\dagger}_{ij} c_i^\dagger c^{\phantom\dagger}_j  +\sum_j \epsilon^{\phantom\dagger}_j c_j^\dagger c^{\phantom\dagger}_j
\end{equation}
with random site energies $\epsilon_j$ drawn from the interval $[-W,W]$ and hopping amplitudes $t_{ij} = e^{i \phi_{ij}}$ for the directed nearest-neighbor bonds $\langle ij\rangle$ with random phases $\phi_{ij}=-\phi_{ji}$. The random phases break time-reversal symmetry, so that the model falls into the unitary symmetry class. Placing the lattice on a torus, we thread the independent loops of the torus by fluxes $\phi_x$ and $\phi_y$. We then compute the corresponding quantum geometric tensor $g_{\alpha\beta}$ by explicitly constructing the current operators, $J_x = \partial_{\phi_x} H$ and $J_y = \partial_{\phi_y} H$ and evaluating the expression in Eq.\ (\ref{eq:Cn}). Figure (\ref{Fig2}) shows the distribution functions of $g_{xx}$ and ${\rm Re} g_{xy} $ for a 3D cubic lattice, where we filter the eigenstates at the center of the band and consider parameters well inside the metallic phase (moderate disorder), such that states at the band center are extended and the elastic mean free path is small compared to the system size $L$. The results are indeed in good agreement with the exact random-matrix distribution. We observe numerically that the off-diagonal elements converge faster to the universal distribution than the diagonal elements. This difference persists for simulations of the corresponding Levy flights and is even more pronounced in simulations of a 2D random flux model. We also confirmed that the Berry curvature ${\rm Im} g_{xy} $ has the same distribution as ${\rm Re} g_{xy}$ in the random flux model. 

{\em Derivation.---}We briefly sketch the derivation of our central result in Eq.\ (\ref{final1}), with details relegated to \cite{supp}. The averages over $H_x$ and $H_y$ in Eq.\ (\ref{Palpha}) reduce to Gaussian integrals and can be readily performed,
\begin{equation}
    \tilde P(\xi_0,\boldsymbol{\xi}) \propto \left\langle \delta(E_N) \prod_{m=1}^{N-1} \frac{E_m^4}{(E_m^2+\frac{i\xi_0}{2N})^2+\frac{|\boldsymbol{\xi}|^2}{4N^2} }             \right\rangle_{H_0}.
\label{inter}
\end{equation}
We reinterpret this as an average over an $(N-1)\times(N-1)$ random matrix $\tilde H$ with eigenvalues $E_m$ and $m=1,\ldots,N-1$, using the joint eigenvalue distribution of the GUE \cite{Oppen1994,Oppen1995}. This yields
\begin{equation}
    \tilde P(\xi_0,\boldsymbol{\xi}) \propto  \left\langle \frac{(\det \tilde H )^6}{\prod_{j=1}^4 \det(\tilde H+ia_j)} \right\rangle_{\tilde H},
\label{basicobject}
\end{equation}
where the parameters $a_j$ with $j=1,\ldots,4$ solve $a_j^2 = i (\xi_0\pm|\boldsymbol\xi|)/2N$.

Equation (\ref{basicobject}) is now amenable to supersymmetry methods (see also Ref.\ \cite{Andreev1995} for a general discussion of spectral determinants in random-matrix theory). One rewrites the determinants as Gaussian integrals over $(N-1)$-dimensional vectors of commuting and anticommuting variables, performs the random-matrix average over $\tilde H$, and employs superbosonization \cite{Bunder2007,Littelmann2008} to reduce the integration over the vectors to a finite-dimensional integral. Computing this integral exactly in the $N\to\infty$ limit by the saddle-point method yields Eqs.\ (\ref{final1}) and \ref{final2}), see \cite{supp} for further details. 

{\em Conclusion}.---We have used supersymmetry techniques to derive the exact distribution function of the quantum geometric tensor for random matrices in the Gaussian Unitary Ensemble and confirmed that it applies to physical models of noninteracting electrons. The matrix elements of the quantum geometric tensor can be thought of as Levy flights with correlations, and some aspects of the resulting distribution resemble corresponding Levy stable distributions. The quantum geometric tensor comprises both the Fubini-Study metric and the Berry curvature. Thus, it plays a central role in semiclassical transport of electrons where it governs gauge forces \cite{Xiao2010}, in the theory of topological phases \cite{Qi2011} where it underlies the definition of Chern numbers, and in the theory of disordered systems where it provides natural scaling variables to understand Anderson localization transitions \cite{Werner2019}. This wide applicability promises numerous applications and extensions of our results to specific physical systems.

{\em Acknowledgement.---}We thank Alex Altland, Christophe Mora, and Miklos Werner for insightful discussions. This work has been supported by CRC 910 of Deutsche Forschungsgemeinschaft, by the National Research, Development and Innovation Office (NKFIH) through the Hungarian Quantum Technology National Excellence Program, project no. 2017-1.2.1- NKP-2017-00001, and by the Fund (TKP2020 IES,Grant No. BME-IE-NAT), under the auspices of the Ministry for Innovation and Technology.

%

\onecolumngrid

\clearpage

\setcounter{figure}{0}
\setcounter{section}{0}
\setcounter{equation}{0}
\renewcommand{\theequation}{S\arabic{equation}}
\renewcommand{\thefigure}{S\arabic{figure}}

\onecolumngrid

\section*{\Large{Supplemental Material}}

\section{Derivation}

\subsection{Quantum geometric tensor}

Inserting a complete set of states into Eq.\ (\ref{QGT}) of the main text, one obtains the expression
\begin{equation}
   g_{\alpha\beta}^{(n)} = \sum_{m(\neq n)} \langle \partial_\alpha \tilde n|\tilde m\rangle\langle \tilde m|\partial_\beta \tilde n\rangle
\label{redsum}
\end{equation}
for the quantum geometric tensor. Differentiating $\langle \tilde m|H|\tilde n\rangle=0$ and using $H_0| m\rangle = E_{m} | m\rangle$ gives 
\begin{equation}
   E_n \langle \partial_\alpha m|n \rangle +E_m \langle  m|\partial_\alpha n \rangle + \langle m| \partial_\alpha H | n\rangle = 0,
\end{equation}
where we specialized to $x=y=0$. Finally using that $\langle \tilde m|\tilde n\rangle =0$ implies $\langle \partial_\alpha m| n\rangle + \langle m|\partial_\alpha n\rangle =0$, one finds 
\begin{equation}
  \langle m|\partial_\alpha n\rangle = \frac{\langle m| \partial_\alpha H|n\rangle}{E_n-E_m}.
\end{equation}
Inserting this into Eq.\ (\ref{redsum}) gives Eq.\ (\ref{eq:Cn}) of the main text. 

\subsection{Integrable systems}

For integrable systems, the eigenvalues $E_n$ can be taken as statistically independent and the spacings $|E_n-E_m|$ in Eq.\ (\ref{eq:Cn}) obey a Poisson distribution. Thus, the distribution $p_s(s)$ of the spacings remains constant in the limit $s\to 0$. A small spacing implies a large term in the sum in Eq.\ (\ref{eq:Cn}). Due to the constant $p_s(s)$ in the limit $s\to 0$, the terms $x\sim 1/s^2$ in the sum in Eq.\ (\ref{eq:Cn}) have a probability distribution $p_x(x)$, which decays at large $x$ as 
\begin{equation}
  p_x(x) = p_s(s)\left|\frac{ds}{dx}\right| \sim \frac{1}{|x|^{3/2}}.
\end{equation}
For this asymptotic decay of $p_x(x)$, both the average and the variance of $x$ diverge. By consequence, in the limit of large $N$, the distribution function of the entire sum in Eq.\ (\ref{eq:Cn}) converges to an appropriate Levy stable distribution with the same asymptotic decay \cite{Bouchaud1990}. The stable distribution depends on whether the signs of the terms in the sum are random (off-diagonal element of the quantum geometric tensor) or not (diagonal element). The characteristic functions of the corresponding stable distributions are given in Eq.\ (\ref{levy}) in the main text. 

We include a heuristic argument yielding Eq.\ (\ref{levy}) for the distribution of the diagonal elements of the quantum geometric tensor. Assuming the existence of a stable distribution, we can choose a convenient distribution $p_x(x)$ for the individual terms in the sum in Eq.\ (\ref{eq:Cn}), with the only requirement that the distribution fall off as $1/|x|^{3/2}$ at large $|x|$. Such a choice is a  Gaussian distribution for the spacings $s$, with the numerators in Eq.\ (\ref{eq:Cn}) simply taken as fixed. As we saw above, the fact that $p_s(s)\sim \exp{-\gamma_0 s^2/4N}$ remains nonzero in the limit $s\to 0$ implies that $p_x(x)\sim 1/|x|^{3/2}$. With this choice, we find
\begin{equation}
   p_x(x) \sim \int_0^\infty ds\, e^{-\frac{\gamma_0}{4N} s^2} \delta(x-\frac{1}{s^2}).
\end{equation}
Here, we focused on the diagonal element of the quantum geometric tensor, for which all terms in the sum in Eq.\ (\ref{eq:Cn}) are positive. We also made the dependence on the matrix size $N$ explicit, choosing the same scalings as for the GUE. Using the Fourier representation of the $\delta$-function, the corresponding characteristic function takes the form
\begin{equation}
  \tilde p_x(\xi) \sim \int_0^\infty ds \exp\left(-\frac{\gamma_0}{4N} s^2-\frac{i\xi}{s^2}\right).
\end{equation}
Here, $\xi$ should be taken to have an infinitesimal negative imaginary part. This integral can be performed and yields
\begin{equation}
   \tilde p_(\xi) = e^{-\sqrt{\frac{\gamma_0}{2N}|\xi|} (1+i {\rm sgn} \xi)}
\end{equation}
Due to statistical independence, the characteristic function $\tilde P(\xi)$ of the entire sum in Eq.\ (\ref{eq:Cn}) is simply given by 
\begin{equation}
  \tilde P(\xi) = [\tilde p(\xi)]^N = e^{-\sqrt{\frac{N\gamma_0}{2}|\xi|} (1+i {\rm sgn} \xi)}.
\end{equation}
This is just a rescaled version of the characteristic function for the distribution of an individual term in Eq.\ (\ref{eq:Cn}) [whose distribution is thus already equal to the Levy stable distribution for our choice of $p_s(s)$] and coincides with Eq.\ (\ref{levy}) in the main text with the identification $\gamma=N\gamma_0$. 

\subsection{GUE average}

Following Refs.\ \cite{Oppen1994,Oppen1995}, we perform the average in Eq.\ (\ref{inter}) of the main text using the joint eigenvalue distribution for $H_0$, 
\begin{equation}
   p_N(E_1,\ldots, E_N) \propto \prod_{i<j}(E_i-E_j)^2 e^{-\frac{1}{2}N \sum_j E_j^2}.
\end{equation}
Writing the terms involving $E_N$ separately and using the large $N$ limit, this is 
\begin{eqnarray}
   p_N(E_1,\ldots, E_N) \propto \prod_{i=1}^{N-1}(E_i-E_N)^2 e^{-\frac{1}{2}N  E_N^2} 
    p_{N-1}(E_1,\ldots,E_{N-1}),
\end{eqnarray}
where $p_{N-1}$ denotes the joint eigenvalue distribution of an $(N-1)\times (N-1)$-dimensional random matrix drawn from the GUE, denoted by $\tilde H$ in the following. Using that the $\delta$-function in Eq.\ (\ref{inter}) allows us to set $E_N=0$, we find 
\begin{equation}
    \tilde P(\xi_0,\boldsymbol{\xi}) \propto \mathbb{E}_{\rm GUE} \left[ \delta(E_N) \prod_{m=1}^{N-1} \frac{E_m^6}{(E_m^2+\frac{i\xi_0}{2N})^2+\frac{|\boldsymbol{\xi}|^2}{4N^2} }\right].
\label{inter2}
\end{equation}
Here, we write the GUE average (denoted by $\langle\ldots\rangle_{\tilde H}$ in the main text) as $\mathbb{E}_{\rm GUE} \left[\ldots \right]$. 

Equation (\ref{inter2}) can be rewritten as a GUE average over determinants of $\tilde H$, as given in Eq.\ (\ref{basicobject}) in the main text. Factorizing the denominator gives 
\begin{equation}
    \tilde P(\xi_0,\boldsymbol{\xi}) \propto \lim_{b_j\to 0}  \mathbb{E}_{\rm GUE} \left[ \frac{\prod_{j=1}^6 \det (\tilde H + \mathrm{i}b_j)}{\prod_{j=1}^4 \det(\tilde H+\mathrm{i}a_j)} \right].
\label{DetExp}
\end{equation}
Here, the $a_j$ with $j=1,\ldots,4$ solve $a_j^2 = i (\xi_0\pm|\boldsymbol\xi|)/2N$. There are two roots with ${\rm Re}\,a_j >0$, which we denote as $a_1$ and $a_3$, and two roots with ${\rm Re}\,a_j <0$, which we denote as $a_2$ and $a_4$. We also introduced parameters $b_j$ with $j=1,\ldots,6$. The $b_j$ need to be set to zero at the end, but it turns out to be convenient to retain them at intermediate steps of the calculation.  

We represent the determinants as Gaussian integrals. The determinants in the denominator are written as integrals over complex variables $z, \bar{z}$ (with Einstein's summation convention in force)
\begin{equation}
    \mathrm{det}^{-1}(\tilde H + \mathrm{i}a)
    = \int_{z,\bar{z}} \mathrm{e}^{\pm \mathrm{i} \bar{z}_k (\tilde H + \mathrm{i}a)^k_{\;\; l} z^l}.
\label{bosgau}
\end{equation}
For convergence, we choose the upper sign when $\mathrm{Re}\, a>0$ and thus for the determinants involving $a_1$ and $a_3$, and the lower sign when $\mathrm{Re}\, a<0$ and thus for $a_2$ and $a_4$. The determinants in the numerator are written as integrals over Grassmann variables $\zeta, \bar\zeta$,
\begin{equation}
    \mathrm{det}[\mathrm{i} (\tilde H+\mathrm{i}b)] = \int_{\zeta,\bar\zeta} \mathrm{e}^{-\mathrm{i} \bar{\zeta}_k (\tilde H+\mathrm{i}b)^k_{\;\; l} \zeta^l} ,
\end{equation}
where we note that $\prod_{j=1}^6\mathrm{det} (\tilde H+\mathrm{i}b_j) = (-1)^N \prod_{j=1}^6\mathrm{det}[\mathrm{i} (\tilde H+\mathrm{i}b_j)] $.

We now collect the random factors into
\begin{equation}
    X \equiv \exp \left\{ \mathrm{i} \tilde H^k_{\;\; l} \left(
    z^{l}_{\; 1} \bar{z}^{1}_{\;\; k} - z^{l}_{\; 2} \bar{z}^{2}_{\;\; k}+z^{l}_{\; 3} \bar{z}^{3}_{\;\; k} - z^{l}_{\; 4} \bar{z}^{4}_{\;\; k}
    + \zeta^{l}_{\; f} \bar\zeta^{f}_{\;\; k}
    \right) \right\} ,
\end{equation}
where $f = 1, \ldots, 6$, and introduce supervectors 
\begin{equation}
    \{ \Psi^{l}_{\; \mu} \} = \left( z^{l}_{\; 1} \,, z^{l}_{\; 2} \,, z^{l}_{\; 3} \,, z^{l}_{\; 4} \,, \zeta^{l}_{\; 1} \,, \zeta^{l}_{\; 2} \,, \zeta^{l}_{\; 3} \,, \zeta^{l}_{\; 4} \,, \zeta^{l}_{\; 5}\,, \zeta^{l}_{\; 6}\right)
\end{equation}
to abbreviate the notation. Then, we have 
\begin{equation}
    \Psi^{l}_{\; \mu} (s \bar\Psi)^{\mu}_{\; k} =
     z^{l}_{\; 1} \bar{z}^{1}_{\;\; k} - z^{l}_{\; 2} \bar{z}^{2}_{\;\; k}+z^{l}_{\; 3} \bar{z}^{3}_{\;\; k} - z^{l}_{\; 4} \bar{z}^{4}_{\;\; k}
    + \zeta^{l}_{\; f} \bar\zeta^{f}_{\;\; k} 
\end{equation}
with 
\begin{equation}
    s = \mathrm{diag}(1,-1,1,-1,1,1,1,1,1,1) .
\end{equation}
Taking the GUE expectation value has now been reduced to a Gaussian integral, which yields
\begin{equation}
    \mathbb{E}_{\rm GUE} (X) = \mathbb{E}_{\rm GUE} \left( \mathrm{e}^{\mathrm{i} \tilde H^k_{\;\; l} (\Psi s \bar\Psi)^l_{\;\; k} } \right) = \mathrm{e}^{(-\lambda^2 / 2N) (\Psi s \bar\Psi)^l_{\;\; k} (\Psi s \bar\Psi)^k_{\;\; l} } \,.
\end{equation}
Using the cyclicity of trace and supertrace, the exponent on the right hand side can be written as a supertrace, 
\begin{equation}
    \mathbb{E}_{\rm GUE} (X) = \mathrm{e}^{- (\lambda^2 / 2N) \, \mathrm{tr} (\Psi s \bar\Psi)^2} = \mathrm{e}^{- (\lambda^2 / 2N)\, \mathrm{STr} (\bar\Psi \Psi s )^2} .
\end{equation}
Here,  $\lambda$ denotes the disorder strength parameter of the GUE, which was set to $\lambda=1$ in the main text.

\subsection{Superbosonization step}

Consider the composite object (with $k = 1, 2, \ldots, N$ for $N\times N$ GUE matrices)
\begin{equation}
    M^{\mu}_{\;\; \nu} = N^{-1} \bar\Psi^{\mu}_{\;\; k} \Psi^{k}_{\;\; \nu} \,.
\end{equation}
This is a supermatrix of dimension $(4|6) \times (4|6)$. The superbosonization method \cite{Bunder2007,Littelmann2008} allows us to switch from the original variables $z,\bar{z}$ and $\zeta,\bar\zeta$ of integration to supermatrices $M$ as new integration variables. In the fermion-boson block decomposition,
\begin{equation}
    M = \left( \begin{matrix} M_{\rm BB} &M_{\rm BF} \cr M_{\rm FB} &M_{\rm FF} \end{matrix} \right) ,
\end{equation}
the block $M_{\rm BB}$ is a positive Hermitian $4 \times 4$ matrix,
\begin{equation}
    M_{\rm BB} = N^{-1} \left( \begin{matrix}
    \bar{z}^{1}_{\;\;k} z^{k}_{\; 1} &\ldots
    &\bar{z}^{1}_{\;\;k} z^{k}_{\; 4} \cr
    \vdots &\ddots &\vdots \cr
    \bar{z}^{4}_{\;\;k} z^{k}_{\; 1} &\dots
    &\bar{z}^{4}_{\;\;k} z^{k}_{\; 4} \end{matrix} \right) ,
\end{equation}
while $M_{\rm FF}$,
 \begin{equation}
    M_{\rm FF} = N^{-1} \left( \begin{matrix}
    \bar{\zeta}^{1}_{\;\;k} \zeta^{k}_{\; 1} &\ldots
    &\bar{\zeta}^{1}_{\;\;k} \zeta^{k}_{\; 6} \cr
    \vdots &\ddots &\vdots \cr
    \bar{\zeta}^{6}_{\;\;k} \zeta^{k}_{\; 1} &\dots
    &\bar{\zeta}^{6}_{\;\;k} \zeta^{k}_{\; 6} \end{matrix} \right) ,
\end{equation}
turns into a unitary $6 \times 6$ matrix, and the entries of $M_{\rm BF}$ and $M_{\rm FB}$ are Grassmann variables. The change of variables is carried out by using the superbosonization identity
\begin{equation}
    \int_{z,\bar{z}} \int_{\zeta,\bar\zeta} F \left( M(\bar{z},z,\bar\zeta,\zeta) \right) =
    \int \mathcal{D} M \, \mathrm{SDet}^N(M) F(M) ,
\end{equation}
where a normalization constant is absorbed into the new integration measure, $\mathcal{D} M$. The new measure is scale invariant and, 
up to a constant,  uniquely determined by the symmetries of the problem. 

\subsection{Saddle-point approximation}

After superbosonization, we have
\begin{equation}
    \tilde P(\xi_0,\boldsymbol{\xi}) = \int \mathcal{D} M \, \mathrm{SDet}^N(M)
    \, \mathrm{e}^{- (N \lambda^2 / 2) \, \mathrm{STr} (M s)^2 - N  \mathrm{STr} (smM)} 
\end{equation}
with $m=\mathrm{diag}(a_1,a_2,a_3, a_4,b_1,\ldots,b_6)$ as defined in the main text. In the limit of large random matrices, $N \to \infty$, the integral can now be performed by saddle-point integration. Since $m \sim N^{-1}$, the corresponding term can be neglected in determining the saddle-point manifold, and the saddle-point equation becomes
\begin{equation}
    M^{-1} - \lambda^2 s M s = 0 \,.
\end{equation}
This has the supermanifold of dominant (for $N\to \infty$) solutions
\begin{equation}
    M s = \lambda^{-1} Q , \quad Q = T \Sigma_3 T^{-1} ,
\end{equation}
where
\begin{equation}
    \Sigma_3 = \mathrm{diag}(1,-1,1,-1,1,-1,1,-1,1,-1) , \quad T \in \mathrm{U}(2,2|6) .
\end{equation}
Thus, saddle-point integration yields
\begin{equation}\label{eq:Ptilde}
   \tilde  P(\xi_0,\boldsymbol{\xi}) = \int DQ\, \mathrm{e}^{- (N/ \lambda)\, {\rm STr} (Qm)} ,
\end{equation}
where $DQ$ is the invariant measure on $\mathrm{U}(2,2|6) / \mathrm{U}(2|3) \times \mathrm{U}(2|3)$.  Up to a multiplicative constant,
this measure is again determined uniquely by symmetries.

\subsection{Semiclassical exactness}

Our integral representation for $\tilde{P}$ is semiclassically exact, c.f., \cite{critique}, which significantly simplifies the calculation. The principle of semiclassical exactness is easiest to apply if the critical points of the integrand are isolated. In the present case, that is not the case once we set $b_j\to0$. It is for this reason that we introduced the $b_j$ at all intermediate stages of the calculation and take the limit $b_j\to0$ only at the very end.  

Now all critical points are isolated and using the semiclassical exactness, we can evaluate the integral (\ref{eq:Ptilde}) semiclassically. The isolated critical points are given by
\begin{equation}
    Q_{\rm crit} = \mathrm{diag}(+1,-1,+1,-1,s_1,s_2,s_3,s_4,s_5,s_6)
\end{equation}
where $s_f \in \{ \pm 1 \}$ and $\sum_f s_f = 0$. There exist $6!/(3! 3!) = 20$ critical points, namely $Q_{\rm crit} = \Sigma_3$ and 19 
more.

Then, the value of the integral (\ref{eq:Ptilde}) is a sum of 20 terms (one for each critical point) and each term contributes by the value of the integral at the critical point times a factor originating from the corresponding fluctuation integral in Gaussian approximation. The contribution from the critical point $Q_{\rm crit} = \Sigma_3$ takes the form
\begin{equation}
    \tilde{P}(\xi_0,\boldsymbol{\xi})_{\Sigma_3} = \frac{\lambda}{N} \, \Delta(\xi_0,\boldsymbol{\xi}) \,
    \mathrm{e}^{- (N / \lambda)\, {\rm STr} (\Sigma_3 m)} ,
\end{equation}
where $\Delta(\xi_0,\boldsymbol{\xi})$ is given by
\begin{equation}
    \Delta(\xi_0,\boldsymbol{\xi}) = \frac{\prod_{i=1}^{2}\prod_{j=2,4,6} (a_i-b_j)\prod_{i=3}^{4}\prod_{j=1,3,5} (a_i-b_j)}{\prod_{i=1}^{2}\prod_{j=3}^{4}(a_i-a_j)\prod_{i=1,3,5}\prod_{j=2,4,6}(b_i-b_j)}.
\label{Delta}
\end{equation}
The contributions from the other 19 critical points $Q_{\rm crit}$ are obtained by applying to $[b_1,b_2,b_3,b_4,b_5,b_6]$ the same permutation that turns $\Sigma_3$ into the given $Q_{\rm crit}$, and $\tilde{P}(\xi_0,\boldsymbol{\xi})$ follows by summing over the contributions of all critical points.

The denominator of Eq.\ (\ref{Delta}) is singular in the limit $b_j\to 0$. However, after summing over all critical points one finds that there is a compensating factor in the numerator and the limit becomes well defined. Performing this calculation \footnote{We have performed this calculation using Mathematica.} gives Eqs.\ (\ref{final1}) and (\ref{final2}).

\begin{figure}[t]
	\includegraphics[width=.25\linewidth]{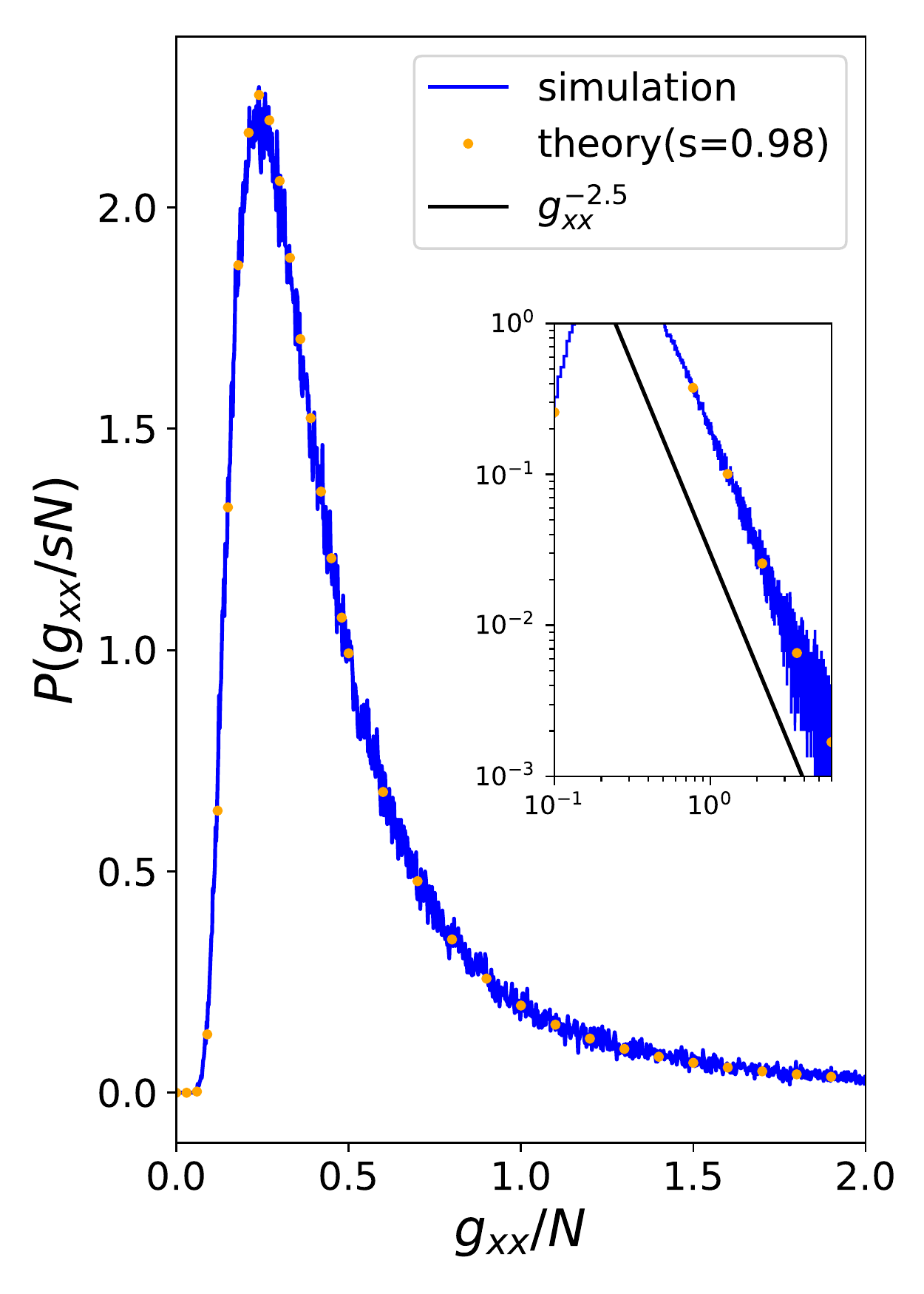} 
	\includegraphics[width=.25\linewidth]{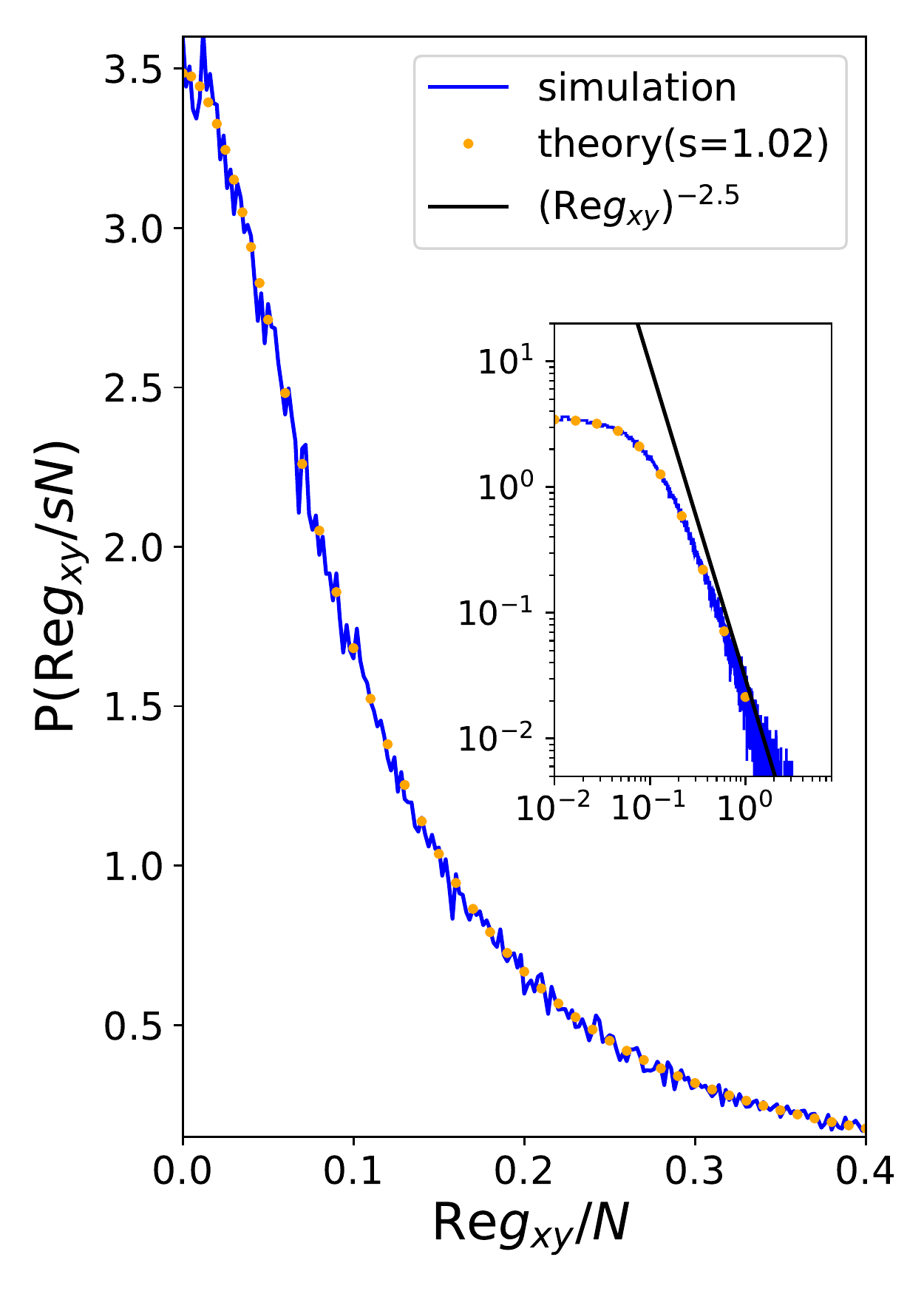}  \\
	\includegraphics[width=.25\linewidth]{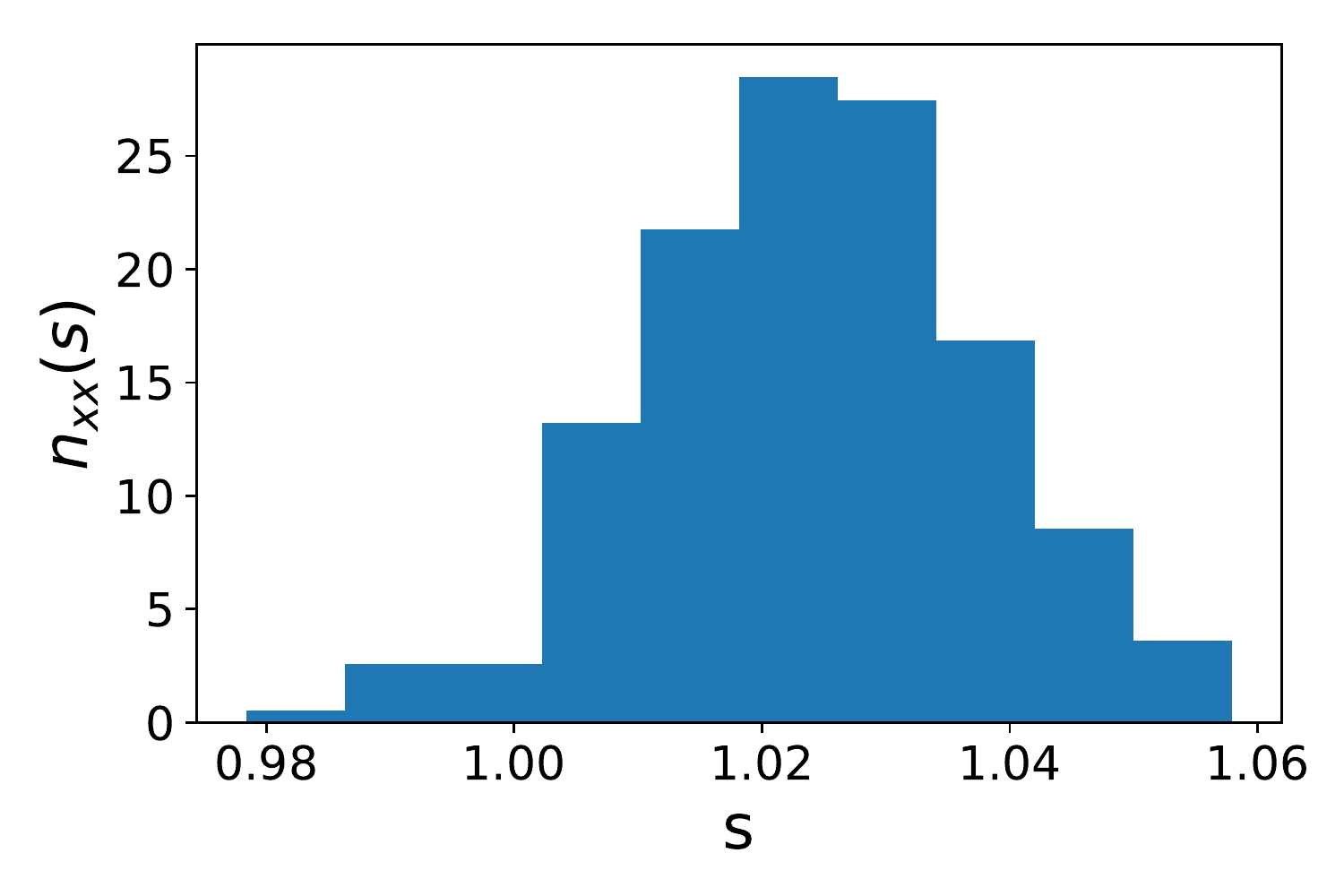} 
	\includegraphics[width=.25\linewidth]{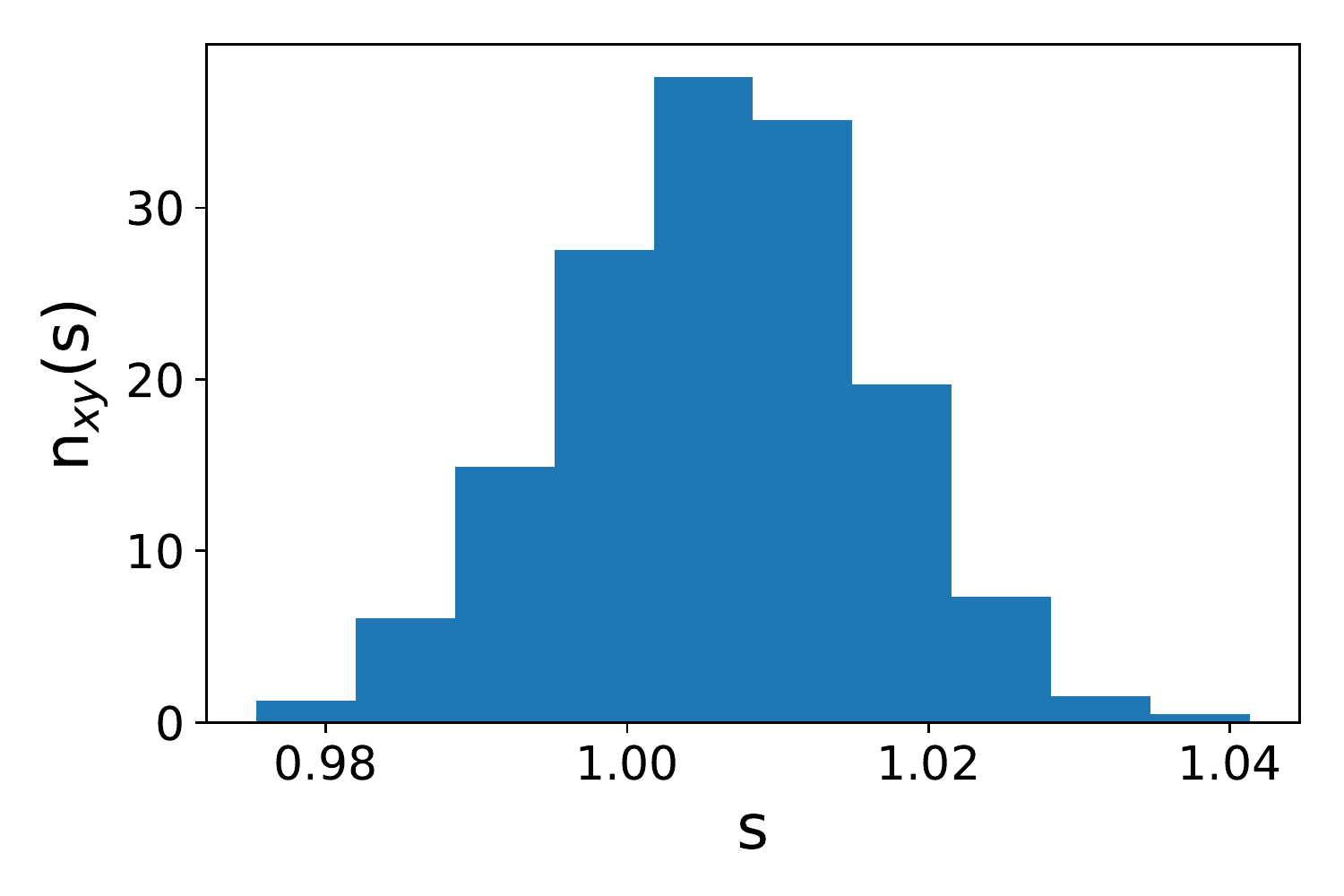} 
		 \caption{\label{SFig1}  Top panels: Distribution functions of matrix elements of the quantum geometric tensor (left: $g_{xx}$; right: ${\mathrm Re} g_{xy}$), obtained by sampling $10^6$ realizations of $H_0$ in Eq.\ (\ref{ham}) with $H_0$ drawn from the Gaussian Unitary Ensemble with $N=100$. The sampling is performed for fixed perturbation matrices $H_x$ and $H_y$ (chosen as matrices drawn independently from the Gaussian Unitary Ensemble). The insets show a corresponding log-log plot, emphasizing the asymptotic $1/|g|^{5/2}$ decay. A plot of $f(g) \propto 1/|g|^{5/2}$ is shown for comparison. Numerical data (blue) are compared to the analytical prediction (orange dots) given in Eq.\ (\ref{final1}) in the main text with $\gamma=s\gamma^\mathrm{GUE}$ and $s=0.978$ (left) and $s=1.018$ (right). Bottom panels: Distribution of scaling factors as defined in Eq.\ (\ref{scalefactors}). The scale factors describe the fits of the distributions of the quantum geometric tensor to our analytical result in Eq.\ (\ref{final1}) and are obtained by sampling and fitting the distributions of $g_{xx}$ and ${\rm Re}g_{xy}$ for 600 sets of random, but fixed perturbation matrices drawn from the GUE. }
\end{figure} 

\section{Averaging over $H_0$ only}

In the main text, we assume that the two parameters $x$ and $y$ couple to independent random matrices, i.e., we average over both the unperturbed Hamiltonian $H_0$ and the perturbations $H_x$ and $H_y$. This assumption can be relaxed. Averaging only over the unpertubed Hamiltonian $H_0$, the matrix elements in the numerator of Eq.\ (\ref{eq:Cn}) are still random variables as they involve the eigenvectors of the GUE matrix $H_0$. In the limit $N\to\infty$, the matrix elements of the perturbation matrices in the eigenbasis of $H_0$ become Gaussian random variables with zero mean and covariance
\begin{eqnarray}
\mathbb{E}_{\rm GUE} \{  \langle n| H_\alpha|m \rangle \langle m| H_\beta | n \rangle  \} &=& \frac{1}{N^2} {\rm tr} H_\alpha H_\beta \\
\mathbb{E}_{\rm GUE} \{  \langle n| H_\alpha|m \rangle \langle n| H_\beta | m \rangle  \} &=& 0.  \,\,\,\,\,\,\,\,\,\,\,\,\,\,\,\,\,\,\,\,\,\,\,\,\,\,\,\,\, (m\neq n)
\end{eqnarray}
As long as we consider perturbations $H_x$ and $H_y$ such that, to leading order in the large-$N$ limit, the covariance matrix 
\begin{equation}
   C_{\alpha\beta} = \frac{1}{N^2} {\rm tr}  H_\alpha  H_\beta
\end{equation}
for $ H_\alpha$ is proportional to the unit matrix, the calculations can now proceed exactly as in the case discussed in the bulk of this paper, in which one averages over the perturbations $H_x$ and $H_y$.  
  
This situation occurs when the perturbations are drawn independently from a GUE, but then held fixed while averaging over $H_0$. The resulting distributions are in excellent agreement with our analytical result. A comparison between the numerical results and the exact distribution of the quantum geometric tensor in Eq.\ (\ref{final1}) is shown in Fig.\ \ref{SFig1} (top panels). The random fluctuations of the strength of the perturbation matrices across the GUE can be accounted for by introducing a scale factor $s$ through
\begin{equation}
     \gamma = s \gamma^\mathrm{GUE},
\label{scalefactors}
\end{equation}
relative to the GUE result $\gamma^{\mathrm{GUE}}=4N$. By fitting the numerical results to Eq.\ (\ref{final1}) for different GUE matrices $H_x$ and $H_y$, we can numerically obtain the corresponding distributions of scaling factors as shown in Fig.\ \ref{SFig1} (bottom panels). In accordance with random-matrix estimates, the deviation of the scale factor from unity is of order $1/N$. 

We note that our approach to computing the joint distribution function for the quantum geometric tensor can also be extended to the case of a general covariance matrix. Then, we first define new perturbations $\overline{H}_\alpha$ and parameters $\overline{\mathbf{r}}=(\overline x, \overline y)$  through
\begin{eqnarray}
  \overline {\boldsymbol{r}} &=& D {\boldsymbol{r}} \\
  \overline H_\alpha &=& \sum_\beta D_{\alpha \beta} H_\beta, 
\end{eqnarray}
where we choose the orthogonal matrix $D$ such that the covariance matrix becomes diagonal. We then have to extend the calculation to situations in which the effective averages over $\overline H_x$ and $\overline H_y$ are still GUE-like, albeit with different disorder parameters $\lambda_x$ and $\lambda_y$. Performing the average over the eigenvectors of the unperturbed Hamiltonian will then result in Eq.\ (\ref{DetExp}) with 
\begin{equation}
   a_j^2 = \frac{i}{4N}\left[ \xi_0(\lambda_x+\lambda_y) + \xi_3(\lambda_x -\lambda_y) \right]  \pm \frac{i}{2N} \sqrt{\frac{1}{4}[\xi_0(\lambda_x-\lambda_y) + \xi_3(\lambda_x +\lambda_y)]^2 +\lambda_x\lambda_y (\xi_1^2 + \xi_2^2)}.
\end{equation}
We first consider the distributions of the diagonal and off-diagonal elements of  the quantum geometric tensor. To obtain the distribution of the off-diagonal elements, we set $\xi_0=\xi_3=0$. In this case, the product $\lambda_x\lambda_y$ simply rescales the otherwise unchanged distribution function. To obtain the distribution functions of  the diagonal elements, we set $\xi_0=\pm \xi_3=\xi$ and $\xi_1=\xi_2=0$. Again, the distribution functions are merely rescaled, though differently for $g_{xx}$ and $g_{yy}$. 
Finally, the joint distribution function follows by setting
\begin{equation}
   \xi_\pm = \frac{1}{2}\left[ \xi_0(\lambda_x+\lambda_y) + \xi_3(\lambda_x -\lambda_y) \right]  \pm \sqrt{\frac{1}{4}[\xi_0(\lambda_x-\lambda_y) + \xi_3(\lambda_x +\lambda_y)]^2 +\lambda_x\lambda_y (\xi_1^2 + \xi_2^2)}
\end{equation}
in the characteristic function in Eq.\ (\ref{final1}), Fourier transforming, and reverting to the quantum geometric tensor with respect to the original parameters $x$ and $y$. 
 
\end{document}